\documentclass[twocolumn,journal]{IEEEtran}
\usepackage[T1]{fontenc}
\usepackage[latin9]{inputenc}
\usepackage{float}
\usepackage{amstext}
\usepackage{amsthm}
\usepackage{amssymb}
\usepackage{graphicx}
\usepackage[unicode=true,
 bookmarks=true,bookmarksnumbered=true,bookmarksopen=true,bookmarksopenlevel=1,
 breaklinks=false,pdfborder={0 0 0},pdfborderstyle={},backref=false,colorlinks=false]
 {hyperref}
\hypersetup{pdftitle={Your Title},
 pdfauthor={Your Name},
 pdfpagelayout=OneColumn, pdfnewwindow=true, pdfstartview=XYZ, plainpages=false}
\usepackage{breakurl}

\makeatletter

\floatstyle{ruled}
\newfloat{algorithm}{tbp}{loa}
\providecommand{\algorithmname}{Algorithm}
\floatname{algorithm}{\protect\algorithmname}

\theoremstyle{plain}
\newtheorem{thm}{\protect\theoremname}

\usepackage[caption=false,font=footnotesize]{subfig}

\makeatother

\providecommand{\theoremname}{Theorem}

\begin{document}

\title{Distributing Complexity: A New Approach to Antenna Selection for
Distributed Massive MIMO}

\author{Harun Siljak, Irene Macaluso, and~Nicola Marchetti\thanks{The authors are with the CONNECT Centre, Trinity College Dublin, Ireland
(emails: \{siljakh, macalusi, marchetn\}@tcd.ie)}\thanks{This publication has emanated from research supported in part by a
research grant from Science Foundation Ireland (SFI) and is co-funded
under the European Regional Development Fund under Grant Number 13/RC/2077.
The project has received funding from the European Union\textquoteright s
Horizon 2020 research and innovation programme under the Marie Sk\l odowska-Curie
grant agreement No 713567.}}
\maketitle
\begin{abstract}
Antenna selection in Massive MIMO (Multiple Input Multiple Output)
communication systems enables reduction of complexity, cost and power
while keeping the channel capacity high and retaining the diversity,
interference reduction, spatial multiplexity and array gains of Massive
MIMO. We investigate the possibility of decentralised antenna selection
both to parallelise the optimisation process and put the environment
awareness to use. Results of experiments with two different power
control rules and varying number of users show that a simple and computationally
inexpensive algorithm can be used in real time. The algorithm we propose
draws its foundations from self-organisation, environment awareness
and randomness. 
\end{abstract}


\IEEEpeerreviewmaketitle{}

\section{Introduction}

\IEEEPARstart{C}{ombining} the classical idea of antennas distributed
in space \cite{kerpez1996radio} and the new concept of massive MIMO
antenna systems \cite{marzetta_noncooperative_2010}, distributed
massive MIMO offers diversity, spatial multiplexing opportunities,
interference suppression and redundance \cite{ozgur2013spatial}.
The question of redundance and the number of antennas needed to operate
in a certain environment can be answered under certain conditions
\cite{hoydis_massive_2013}, and it reinforces the importance of antenna
selection. Antenna selection in Massive MIMO system can help with
power optimisation, complexity reduction and provide a set of antennas
available for other purposes, e.g. nulling \cite{geraci_operating_2017}.
Of course, a trade-off between these benefits exists \cite{liu_energy_2017}
and different objectives and performance measures are used, from spectral
efficiency and constructive interference \cite{amadori_interference-driven_2016}
to fairness measures \cite{makki_genetic_2017}. Other factors are
also taken into account, such as the circuit power consumption \cite{hamdi_resource_2016}.
Another aspect investigated is the difference between FDD (frequency
division duplex) and TDD (time division duplex) in terms of CSI (channel
state information) collection, which is essential for antenna selection
\cite{benmimoune_novel_2017}. These algorithms are most often centralised
and based on co-located antenna systems.

A large number of antenna selection algorithms for MIMO (and recently
massive MIMO) has been proposed over the last 15 years, one of the
first being the removal of antennas highly correlated with other antennas
in the selected set \cite{molisch_capacity_2005}. Several approaches
were based on the greedy principle: in \cite{gharavi-alkhansari_fast_2004},
the authors proposed an iterative algorithm that starts from an empty
set of selected antennas and in each turn picks the antenna that contributes
the most to the capacity of the selected antennas set. Its dual, i.e.
the algorithm that starts with all the antennas and removes those
that contribute the least iteratively, has been presented in \cite{gorokhov_receive_2003}.
Both algorithms terminate once the desired number of selected antennas
is reached. A case for environment-ignoring random selection was made
as well, pointing out that for a large number of selected antennas
it is comparable to other selection procedures \cite{lee_energy_2013},
which has been observed in practice for planar co-located massive
MIMO \cite{gao_massive_2015}. To exemplify the various approaches
used, we note that methods using convex optimisation \cite{mahboob_transmit_2012},
combinatorial optimisation \cite{michalopoulos_distributed_2008},
genetic algorithm \cite{makki_genetic_2017} have been proposed.

In this letter we propose a novel distributed, local environment-aware
antenna selection algorithm based on sum-capacity maximisation. The
aim is to achieve sum rates comparable to those achievable through
the use of more computationally complex centralised algorithms and
allow flexibility and adaptability of the scheme. By distributing
the computation over the nodes, we reduce computational complexity
and allow the systemic complexity to enhance the performance.

\section{The Antenna Selection Algorithm}

We consider the scenario of downlink (transmit) antenna selection
at the distributed massive MIMO base station with $N_{T}$ antennas.
In the cell there are $N_{R}$ single antenna users and we aim at
maximising the sum-capacity

\begin{equation}
\mathcal{C}=\max_{\mathbf{P},\mathbf{H_{c}}}\log_{2}\det\left(\mathbf{I}+f(\rho,\,N_{TS},\,N_{R})\mathbf{H_{c}}\mathbf{P}\mathbf{H_{c}}^{H}\right)
\end{equation}
where $\mathbf{I}$ is $N_{TS}\times N_{TS}$ identity matrix, $\mathbf{P}$
is a diagonal $N_{R}\times N_{R}$ matrix describing the power distribution
and $\mathbf{H_{c}}$ is the $N_{TS}\times N_{R}$ channel matrix
representing a selected subset of antennas from a set of $N_{T}$
antennas ($N_{T}\ge N_{TS}$) represented in the channel matrix $\mathbf{H}$,
sized $N_{T}\times N_{R}$ \cite{gao_massive_2015}. The term $f(\rho,\,N_{TS},\,N_{R})$
represents the transmission power factor from the downlink channel
model
\begin{equation}
\mathbf{y}=\sqrt{f(\rho,\,N_{TS},\,N_{R})}\mathbf{H_{c}z}+\mathbf{n}
\end{equation}
with $\mathbf{y}$ being the $N_{R}\times1$ received vector, $\mathbf{z}$
being the $N_{T}\times1$ transmit vector and $\mathbf{n}$ representing
the noise vector; $\rho$ is the signal to noise ratio (SNR) at each
user.

There are two ways the power can be managed in downlink, and we address
them both. One is harvesting the array gain by improving the SNR at
the user side by taking $f(\rho,\,N_{TS},\,N_{R})=\rho N_{R}$ (power
control A), while the other harvests the array gain as a means of
reducing transmit power, achieved through taking $f(\rho,\,N_{TS},\:N_{R})=\rho N_{R}/N_{TS}$
(power control B). In both cases, the increase in the number of users
increases the transmit power. These two optimisation problems are
fundamentally different. The problem using the power control A is
akin to the receiver antenna selection. This problem can be solved
using greedy algorithms with a guaranteed (suboptimal) performance
bound, simply adding antennas in an initially empty set of selected
antennas based on the contribution to the sum rate they bring in.
The problem with model B is that it is not submodular (in particular,
not monotonic) \cite{vaze_sub-modularity_2012}. This means that the
addition of an antenna in the selected set of antennas can decrease
channel capacity, and greedy algorithms cannot provide performance
guarantees.

The optimisation problem we are solving is twofold. We are looking
for both the subset of the total set of available antennas and for
the optimal power distribution over them. Following the practice from
\cite{gao_massive_2015}, we initially assume all diagonal elements
of $\mathbf{P}$ equal to $1/N_{R}$ (their sum is unity, making the
total power equal to $f(\rho,\,N_{TS},\,N_{R})$), perform the antenna
selection and then perform the selection of matrix $\mathbf{P}$ using
water filling for zero forcing. The choice of zero forcing for precoding
was a matter of practicality, the antenna selection algorithm we propose
is independent of the channel model or the precoding scheme.

In the following explanation of the algorithm we will use the term
neighbourhood for the set of $k$ elements/antenna indices denoting
the $k$ nearest neighbours of an antenna ($D_{i}$: the $i$th antenna's
neighbourhood). Neighbourhood of an antenna $A$ is shown in Fig.
\ref{antennas}. Flag bit $f_{j}$ represents the on/off state our
algorithm proposes for the $j$th antenna in an iteration.

Our local algorithm is motivated by a simple model in which every
antenna communicates with its neighbourhood to determine whether it
should participate in the set of selected antennas $S_{i}$ from the
neighbourhood $D_{i}$ or not, i.e. $S_{i}=\{j\in D_{i}|f_{j}=1\}$.
Each antenna element node calculates the sum-capacity with power model
B for the currently selected set of antennas from its neighbourhood,
and then for this set augmented with its antenna.\footnote{The calculation of local sum capacity using power control B is important
for the algorithm as the case of control A would allow every antenna
to join and improve the selected set by just adding more power to
it. Control B allows only the antennas improving the information content
to join in.} If the latter is larger than the former, the node sets its flag to
one (else, it resets to zero). 

Starting from a random selection of antennas, this simple rule in
principle organises the antennas either into a stable configuration
or an oscillation between two configurations. As these state(s) may
be a local but not a global maximum of the achievable sum rates, we
introduce a mutation flipping every ($1/p_{M}$)th flag on average.
In case of long coherence intervals, the mutation happens often as
the algorithm gets a chance to run longer on the same CSI. In our
case, it was a rare event as we assumed a short coherence period.\footnote{While the mutation bears resemblance to the genetic algorithm approach
in \cite{makki_genetic_2017}, the core process is different. In our
algorithm, we converge to the best antenna selection by a simple but
\emph{directed} search strategy. In addition to that, we use the \emph{local}
measure of the capacity as the selection criterion.}

The algorithm runs on the same CSI for a prescribed number of iterations
$N_{i}$. After the last iteration, antennas turn on and off to form
the configuration from the iteration that had the highest total capacity
over all antennas. Changing the physical state of the antennas after
each iteration is inefficient: this is why we work with flag bits.
Size of $N_{i}$ depends on the coherence interval.

Algorithm 1 presents the self-organising behaviour we described. We
dub the algorithm \emph{local} to emphasise the locality of computation
and perspective.

\begin{algorithm}[tb]
{\footnotesize{}\caption{The proposed local antenna selection procedure}
}{\footnotesize \par}
\begin{enumerate}
\item {\footnotesize{}Start from a random seed of $n$ flags ``on'', the
rest being ``off''.}{\footnotesize \par}
\item {\footnotesize{}For each antenna $i$, $1\le i\le N_{T}$ perform
the following:}{\footnotesize \par}
\begin{enumerate}
\item {\footnotesize{}Calculate the sum-capacity $\mathcal{C}_{i-}$ of
the system with the channel matrix $H_{S_{i}}$ (the antennas in $S_{i}$)}{\footnotesize \par}
\item {\footnotesize{}Calculate the sum-capacity $\mathcal{C}_{i+}$ of
the system $H_{S_{i}\cup i}$}{\footnotesize \par}
\item {\footnotesize{}Compare the two values: if $\mathcal{C}_{i+}>\mathcal{C}_{i-}$,
the antenna in question should be selected (i.e. on). Otherwise it
should be off. }{\footnotesize \par}
\end{enumerate}
\item {\footnotesize{}Update the flags $f_{i}$ of each of the $N_{T}$
antennas to the result of 2(c) or its opposite, if a random mutation
occurs (with a probability of $p_{M}$). Store the current configuration.}{\footnotesize \par}
\item {\footnotesize{}Repeat $N_{i}$ times steps 2 and 3 before updating
the physical state of the antennas (on and off) to the flag state
that resulted in the highest total capacity.}{\footnotesize \par}
\item {\footnotesize{}Repeat steps 2-4.}{\footnotesize \par}
\end{enumerate}
\end{algorithm}

\section{Sum rates: Local vs. Greedy Algorithm}

\begin{figure}[tb]
\begin{centering}
\includegraphics[width=1\columnwidth]{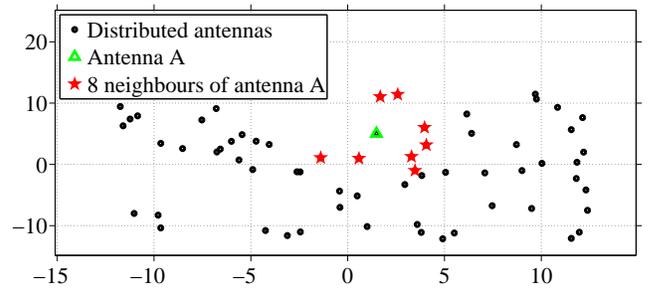}
\par\end{centering}
\vspace{-5mm}

\caption{64 distributed transmitters in the area which included 75 randomly
distributed scatterers and one large obstacle. The number of (randomly
distributed) users varied from 4 to 16. We used 300 OFDM subcarriers,
SNR $\rho=-5\ \text{dB}$, 2.6 GHz carrier frequency, 20 MHz bandwidth
and omnidirectional antennas for transmitters and receivers.}

\vspace{-5mm}
\label{antennas}
\end{figure}

The algorithm was tested using raytracing Matlab tool Ilmprop \cite{galdo2005geometry}
on a system composed by 64 antennas randomly distributed in space
at the same height and shown in Fig. \ref{antennas}. In all computations,
CSI was normalised to unit average energy over all antennas, users
and subcarriers \cite{gao_massive_2015}.

\begin{figure}[tb]
\begin{centering}
\includegraphics[width=1\columnwidth]{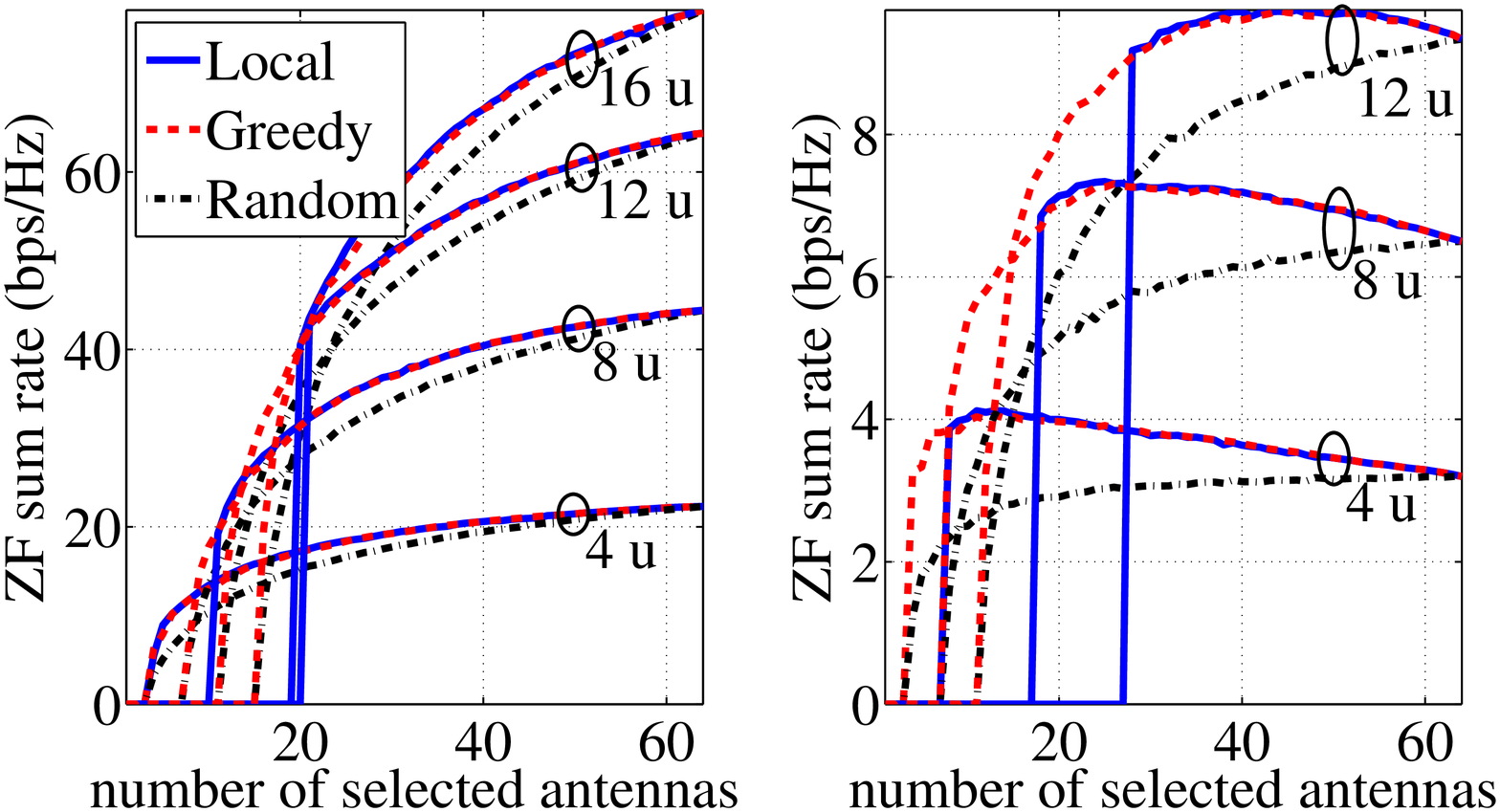}
\par\end{centering}
\vspace{-2mm}

\caption{Antenna selection effects on ZF sum rate for different number of users
in scenarios with power control A and B}

\vspace{-3mm}
\label{locvsrand}
\end{figure}

The two power control scenarios described before are tested and the
results are shown in Fig. \ref{locvsrand} (16 users case omitted
for clarity in the second scenario plot). A clear difference between
the results of local antenna selection and random selection of antennas
was expected: while such a difference is often not detectable in the
case of planar co-located massive MIMO, the distributed case is closer
to cylindrical arrays in this sense, antennas being more far apart
\cite{gao_massive_2015}. The curve representing the local antenna
selection shows the highest sum rate values obtained for different
numbers of selected antennas. It is compared with the sum rates obtained
for the average random case of choosing the same numbers of antennas
and the results of the previously described greedy algorithm (the
one starting with an empty set of antennas \cite{gharavi-alkhansari_fast_2004}).
The sum rates obtained through the proposed local algorithm are on
par (even marginally better) as those obtained through the greedy
algorithm (very low antenna counts for which our algorithm gives zero
rates are not relevant, as they are within $N_{TS}\lesssim N_{R}$
range, not enabling proper beamforming for all users). The proposed
local selection mechanism achieves a comparable performance to the
greedy selection, while reducing the computational complexity.

\begin{figure}[tb]
\begin{centering}
\includegraphics[width=1\columnwidth]{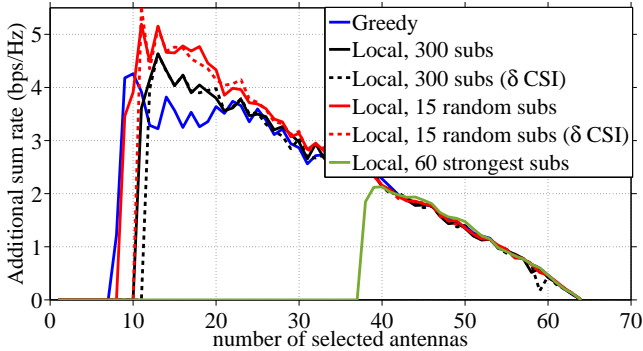}
\par\end{centering}
\vspace{-3mm}

\caption{ZF sum rate for different antenna selection algorithms in 8 users
case with power control A represented as the additional sum rate with
respect to random selection. $\delta$CSI: variation of 30\% in imperfect
CSI measurement.}

\vspace{-5mm}

\label{12comp}
\end{figure}

\section{Computational and Systemic Complexity}

\subsection{Computational Complexity}
\begin{thm}
The worst case complexity of the proposed algorithm is $\mathcal{O}(N_{T}^{\omega})$,
where $\omega$, $2<\omega<3$ is the exponent in the employed matrix
multiplication algorithm complexity.\footnote{It is not always feasible to use the multiplication algorithm with
the lowest complexity due to large constant factors making it hard
to do even a single iteration: the current lowest complexity algorithms
($\omega=2.373$) cannot be implemented in technology at all.}.
\end{thm}
\begin{IEEEproof}
We first note that $\det(\mathbf{I}+\mathbf{AB})=\det(\mathbf{I}+\mathbf{BA})$
for properly sized identity matrices $\mathbf{I}$ on both sides (Sylvester
identity), so we can shift from $N_{T}\times N_{T}$ matrix $\mathbf{H}\mathbf{H}^{H}$
to a smaller $N_{R}\times N_{R}$ matrix $\mathbf{H}^{H}\mathbf{H}$.
The total number of matrix multiplications is $2N_{i}$ and they would
be conducted using some of the standard algorithms with complexity
$\mathcal{O}(N_{R}^{\omega-1}N_{T})$, $2<\omega<3$ \cite{knight1995fast}\footnote{If two $n\times n$ square matrices are multiplied with complexity$\mathcal{O}(n^{\omega}),$
rectangular matrices $a\times b$ and $b\times c$ are multiplied
with complexity $\mathcal{O}(n_{1}^{\omega-2}n_{2}n_{3})$ where $n_{1}=\min(a,b,c)$
and $n_{1},\ n_{2}$ are the other two dimensions. }. This brings the multiplication complexity of our algorithm to $\mathcal{O}(N_{i}N_{R}^{\omega-1}N_{T})$.
In the worst case $N_{R}=N_{T}$, and since $N_{i}=\text{const}$
we obtain the complexity of $\mathcal{O}(N_{T}^{\omega})$. The number
of determinant calculations is constant ($\mathcal{O}(N_{i})$), so
the overall complexity is $\mathcal{O}(N_{T}^{\omega})$.
\end{IEEEproof}
Comparing this to the case of the greedy algorithms with complexity
$\mathcal{O}(N_{T}^{2}N_{R}^{2})$ \cite{gorokhov_receive_2003} and
$\mathcal{O}(N_{T}N_{R}N_{TS})$ \cite{gharavi-alkhansari_fast_2004},
we see that the complexity is reduced: the effect is best seen for
a large number of antennas (massive MIMO) where the constant nature
of $N_{i}$ enables efficient scaling. We also note that $N_{R}$
and $N_{T}$ in our algorithm consideration are not the entire set
of transmitters and receivers as in the greedy algorithms, but just
a $k$-neighbourhood of transmitters and the receivers in their vicinity.
Hence, the computational complexity decreases even more in the practical
implementation.

The main computational burden of the single iteration of both algorithms
is the fact that it is repeated for each of the $c$ OFDM subcarriers.
A natural question to ask is whether we need to optimise over the
whole set of subcarriers, and if a significantly smaller subset could
be selected to represent the channel appropriately. In our study we
propose a random selection of a subcarrier subset and argue that 5\%
of the whole set is enough for practical use, based on the results.
This approach gives a good representation of the channel for the algorithm
as the procedure is repeated $N_{i}$ times, allowing most of the
subcarriers to appear in different iterations and influence the antenna
selection. The alternative is selecting a fixed number of the subcarriers
with the largest average power over all users. Using any subset of
subcarriers could also speed up the greedy algorithm, but our local
selection algorithm still has lower complexity. Fig. \ref{12comp}
represents the gain of sum capacity over random antenna selection
for eight users scenario where we compare the results of greedy selection
and three local algorithm variants: the original proposal using all
300 subcarriers, one with 15 randomly selected subcarriers and one
with 60 strongest subcarriers selected. The random subcarrier selection
variant is computationally superior to both alternatives. It is marginally
better in terms of sum rates and we can observe from Fig. \ref{12comp}
that its minimum selection sets are in general smaller than those
of the full 300 subcarrier algorithm. The strongest subset-based variant
a limited applicability due to a large minimum number of selected
antennas.

Another issue with the growing complexity is the collection of CSI.
It is unrealistic to expect perfect knowledge of CSI, but as Fig.
\ref{12comp} shows, it is not necessary: a 30\% uncertainty in CSI
hs negligible impact on the result of antenna selection.

\begin{figure}[tb]
\begin{centering}
\includegraphics[width=1\columnwidth]{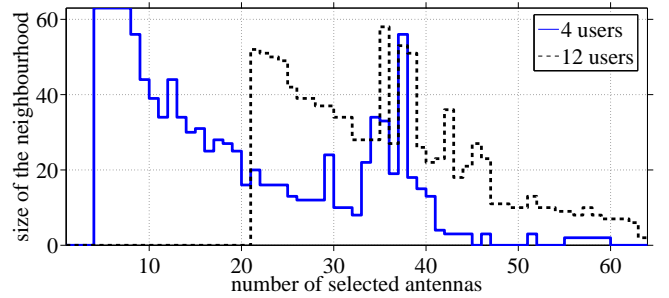}
\par\end{centering}
\caption{The relationship between the number of selected antennas and the number
of neighbours taken in the algorithm}

\vspace{-7mm}
\label{neighsize}
\end{figure}

\subsection{Systemic Complexity: Self-Organisation}

Our algorithm does not start with a predefined number of antennas
to be selected, but converges to a subset whose size depends on the
size of the neighbourhood. This means that for the comparison with
the greedy algorithm shown in Figs. \ref{locvsrand} and \ref{12comp}
we have run the algorithm for varying numbers of neighbours considered.
Fig. \ref{neighsize} shows the size of the neighbourhood observed
and the number of selected antennas in each of those cases for scenarios
with 4 and 12 users (the other two scenarios omitted for clarity)
in the power control scheme A. The variable parameter is placed on
the ordinate axis to align the graph with Fig. \ref{locvsrand}.

The figure demonstrates that the smallest number of antennas is selected
in the case when a single antenna sees most of the other antennas
as its neighbours, and vice versa, the largest number of antennas
is selected in correspondence to the smallest neighbourhoods. Small
neighbourhoods make large effort to support all users, hence turning
on most of their antennas, leading to a high total count of selected
antennas. In large neighbourhoods suboptimal antennas keep themselves
out of the selected subset, seeing the better antennas already in.
We also note the characteristic bimodal shape, implying that in large
neighbourhoods the algorithm switches (oscillates) between small number
of selected antennas and roughly 50\% of the total number. This is
a consequence of the power control B we use (cf. the location of maxima
of the sum rates in Fig. \ref{locvsrand}(b)) as roughly half of the
antennas do not contribute anything new once the other antennas are
included in the selected set. In large neighbourhoods, entropy is
low as all antennas know how they fare against other antennas, making
them more aware of the environment and the rest of the system.

Figs. \ref{locvsrand} and \ref{neighsize} also show that the minimal
number of selected antennas is not always the number of users: the
algorithm sometimes adds more antennas for better beamforming.

\section{Conclusions}

We have presented a novel local antenna selection method for distributed
massive MIMO. Its local nature allows it to be environment-aware and
enables distributed computing at every node. While reducing the complexity
of matrix operations and distributing it over all antenna nodes, we
have additionaly reduced computational complexity by using a very
small subset of subcarriers for optimisation, reducing the time cost
by 20 times. This reduction resulted in both enabling the real-time
application of the algorithm in dynamic environments with short coherence
time and retaining sum-rates on par with other antenna selection algorithms.

Relying on self-organisation, this algorithm emphasises the local
properties of distributed massive MIMO and supports its modularity,
namely the option of cluster separation and distributed control. The
distributed control aspect allows user selection aided by our algorithm
for users close to a neighbourhood cluster, and also the service consolidation
in case of device failures within a cluster. The neighbouring antennas
are aware of local faults and organise themselves accordingly in an
emergent manner, building up on the inherent systemic complexity of
the antenna selection algorithm. The clusters may operate on their
own and/or interact with other clusters, depending on the set neighbourhood
size.

The local selection algorithm relies on randomness in two ways. Randomly
selecting subcarriers to do the optimisation on, it keeps the diversity
of the full subcarrier set while reducing computation time. Randomly
performing mutations on state transitions, it allows leaving local
maxima.

The reduced computational complexity and environment awareness enable
a flexible real time application. Being independent from precoding
choice, channel model and the form of power control, the algorithm
has been shown to perform well in two different variants of transmit
antenna selection.

\bibliographystyle{unsrt}

\end{document}